\documentstyle[psfig]{mn}

\title[Optical identification of a large flat spectrum
sample using APM]{Automated optical identification of a large complete 
northern hemisphere sample of flat spectrum radio sources 
with $S_{\rm6cm}>200$ mJy}

\author[I. Snellen et al.]{I.A.G. Snellen$^{1,2}$, R.G. McMahon$^{1}$, I.M. Hook$^2$, I.W.A. Browne$^3$\\ 
$^1$ Institute of Astronomy, Madingley Road, Cambridge CB3 0HA, United Kingdom\\
$^2$Institute for Astronomy, University of Edinburgh Royal Observatory, Blackford Hill, Edinburgh EH9 3HJ, UK\\ 
$^3$ NRAL Jodrell Bank, University of Manchester, Macclesfield, Chesire SK11 9DL, UK\\}

\date{}

\begin{document}
\maketitle

\begin{abstract}
This paper describes the automated optical APM identification
of radio sources from the Jodrell Bank - VLA Astrometric Survey (JVAS), 
as used for the search for distant radio-loud quasars.
Since JVAS was not intended to be complete, a new complete 
sample, JVAS++, has been constructed 
with selection criteria similar to those of JVAS ($S_{5GHz}>200$ mJy, 
$\alpha_{1.4-5GHz}>-0.5$), and with the use of the more 
accurate GB6 and NVSS surveys. Comparison between this sample and 
JVAS indicates that the completeness and reliability
of the JVAS survey are $\sim 90\%$ and $\sim 70\%$ respectively.
The complete sample has been used to investigate possible 
relations between optical and radio properties of flat spectrum radio 
sources. From the 915 sources in the sample, 756 have 
an optical APM identification at a red ($e$) and/or blue ($o$) plate,
resulting in an identification fraction of 83\% with a completeness and 
reliability of 98\% and 99\% respectively. About $20$\% are optically 
identified with extended APM objects on the red plates, e.g. galaxies. 
However the distinction 
between galaxies and quasars can not be done properly near the magnitude 
limit of the POSS-I plates. 
The identification fraction appears 
to decrease from $>$90\% for sources with a 5 GHz flux density of $>1 Jy$, to
$<$80\% for sources at 0.2 Jy. The identification fraction, in particular that
for unresolved quasars, is found to be lower for sources with steeper radio 
spectra. In agreement with previous studies, we find that the
quasars at low radio flux density levels also tend to have fainter
optical magnitudes, although there is a large spread. 
In addition, objects with a steep radio-to-optical spectral index 
are found to be mainly highly polarised quasars, supporting the 
idea that in these objects the polarised synchrotron component
is more prominent. It is shown that the large spread
in radio-to-optical spectral index is possibly caused by 
source to source variations in the Doppler boosting of the 
synchrotron component.
\end{abstract}

\section{Introduction}

Over the last decade, a comprehensive catalogue of compact flat spectrum
radio sources, the Jodrell Bank $-$ VLA Astrometric Survey (JVAS), has
been constructed (Patnaik et al, 1992; Browne et al., 1997; Wilkinson et al., 
1998). The main aim of this survey was to provide the astronomical community
with a network of bright radio sources with accurate positions 
($\sim 10-15$ mas) primarily intended for use as phase calibrators 
for the Jodrell Bank MERLIN, the VLA and VLBI networks.
In addition to its astrometric goals, JVAS proved an effective means
of finding radio sources exhibiting strong gravitational lensing, with
six systems confirmed to date (King et al. 1999).

Its virtually complete northern-sky coverage makes JVAS uniquely suited
for studying the high luminosity end of the flat spectrum radio source 
population. 
Our group has been particularly involved in the search for distant quasars 
(Hook et al. 1996, Hook et al. 1998, Hook \& McMahon 1998). 
As part of this project, sources in the JVAS survey were optically identified
using the catalogue output of the APM 
(the Automated Plate Measurement Facility at Cambridge) scans of the POSS-I 
plates. This catalogue is ideal for identifying large samples of objects,
since they are generated in an automated 
way allowing an objective assessment of their reliability and homogeneity.

This paper describes the automated optical identification procedure of JVAS,
as used for the search for distant quasars.
It can be used as a guidance for future projects, involving the 
optical identification of comprehensive samples
of flat spectrum radio sources, like those in the Cosmic Lens
All Sky Survey (CLASS; Myers et al. 2001).
Since JVAS is not complete, a new sample is constructed
using selection criteria similar to those of JVAS, but 
with the emphasis on completeness, and with the use of more accurate
selection surveys. By comparing this sample with JVAS, the 
completeness of JVAS is assessed. This is described in section
2. Section 3 and 4 describe the APM-POSS-I catalogue and the optical
identification procedure of JVAS and its complete counterpart. 
Section 5 gives the results and discusses
possible relations between optical and radio properties using the complete
sample.

\section{A complete sample of flat spectrum radio sources}
\subsection{The Jodrell Bank $-$ VLA Astrometric Survey (JVAS)}

The JVAS catalogue has been presented in three separate papers, 
for the regions $+35^\circ\le\delta\le+75^\circ$ (Patnaik et al., 1992),
$+0^\circ\le\delta\le+20^\circ$ (Browne et al., 1997), and 
$+20^\circ\le\delta\le+35^\circ$ and $+75^\circ\le\delta\le+90^\circ$
(Wilkinson et al., 1998). Their initial source list  was constructed using
the Greenbank surveys conducted at 1.4 and 5 GHz by Condon \& Broderick 
(1985, 1986) and Condon, Broderick \& Seielstad (1989), selecting 
all sources with spectral indices larger than $\alpha = -0.5$
($\alpha$ defined as $S_\nu \propto \nu^\alpha$) and 
$S_{5 GHz} \ge 200$ mJy, excluding the 
region $|b|<2^\circ.5$. 
The 5 GHz flux densities used to construct the sample were determined 
directly from the maps and have been found to be systematically 
higher by $\sim 10\%$ than the flux densities determined by
Gregory and Condon (87GB, 1991) from the same maps.
Since the main goal of JVAS was to construct a 
grid of phase calibrator sources, and the distribution of the sample
above contained some undesirable large holes ($\sim 5^\circ$) in
some areas, a few additional sources were selected 
with $S_{5 GHz} \ge 150$ mJy to fill these holes.
Where possible, the potential calibrator sources were cross-checked
against lower frequency surveys to make sure they had genuine flat spectra.
The resulting source list was observed with the VLA at 8.4 GHz between 1990
and 1992 (Patnaik et al, 1992; Browne et al., 1997; Wilkinson et al., 1998).
The resulting catalogue includes sources with observed 8.4 GHz peak brightness
$\ge 50$ mJy/beam. 

For sources that are in the regions $+0^\circ\le\delta\le+20^\circ$ and
$+35^\circ\le\delta\le+75^\circ$ the rms position error was estimated to
be approximately 10 mas in each coordinate 
(Patnaik et al., 1992; Browne et al., 1997). For sources in the regions
$+20^\circ\le\delta\le+35^\circ$ and $+75^\circ\le\delta\le+90^\circ$ this
was estimated to approximately be 40 mas (Wilkinson et al., 1998).
The resulting catalogue contains 2121 radio sources. 

\subsection{The selection of a complete flat spectrum sample}

\begin{table}
\begin{tabular}{ccrrlc}
Survey&Freq.&Wave- &Reso-     &Flux   &Position\\
      &     &length&lution    &Density&Error   \\
      &(GHz)& (cm)   & (arcsec)   &  (mJy)  &(arcsec)  \\
87GB  & 4.85& 6.2  &  210     &$>$25  &10$-$30 \\
GB6   & 4.85& 6.2  &  210     &$>$18  &10$-$30 \\
GB1400& 1.40&21.4  &  710     &$>$150 &30$-$70 \\
NVSS  & 1.40&21.4  &   45     &$>$2.5 &1$-$7   \\
\end{tabular}

\caption{\label{radiosurveys} The relevant parameters of the radio surveys
used for the selection of JVAS and JVAS++.}
\end{table}

The JVAS catalogue was primarily intended to provide phase calibrators with a 
uniform sky distribution, and was not intended as a statistically complete
flat spectrum sample.
Therefore, a similar, complete sample has to be defined 
 to be able to conduct statistically meaningful optical/radio 
studies. This sample, which we call JVAS++, will then also be used to 
assess the completeness of JVAS.

\begin{figure}
\psfig{figure=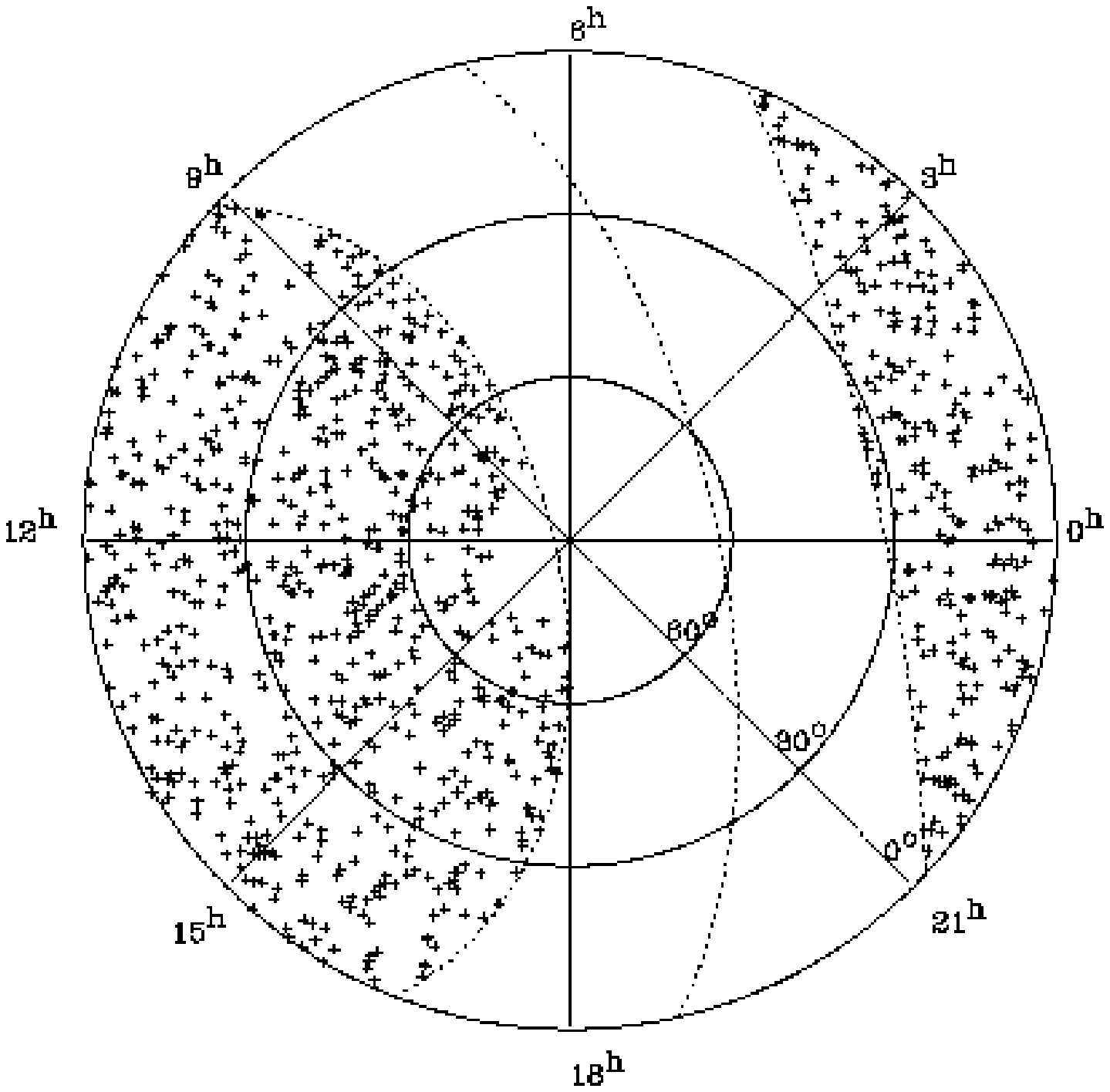,width=8.5cm}
\caption{Polar projection of the sky 
distribution of the sources in the complete sample. The dotted lines 
indicate the $0^\circ$ and $\pm 30^\circ$ galactic latitudes.
\label{skyplot}}
\end{figure}

The JVAS++ sample was constructed using similar selection criteria 
as used for the Cosmic Lens All Sky Survey (CLASS; Myers et al. 2001), and
is based on the recently available and more accurate
radio surveys, the GB6 at 5 GHz (Gregory et al. 1996), and the NVSS at 1.4 GHz 
(Condon et al. 1998). All the relevant parameters of the radio surveys
involved are given in table \ref{radiosurveys}.
In the selection procedure, the 1.4 GHz flux density of a GB6 radio source
is defined as all the NVSS flux within 70$''$ of the GB6 position (as used
for CLASS).
 The sample has the following selection criteria:
\begin{itemize}
\item[1] GB6 5 GHz flux density, $S_{5GHz}>200$ mJy
\item[2] declination, $0^\circ<\delta<75^\circ$, galactic latitude, 
$|b|>30^\circ$.
\item[3] radio spectral index, $\alpha_{1.4-5GHz}>-0.5$.
\end{itemize}
In this way, 915 sources were selected. The distribution on the sky is 
shown in figure \ref{skyplot}. The high galactic latitude 
cut-off was chosen to reduce the possible influence of 
galactic foreground extinction. Furthermore, the APM catalogue is 
only available at these latitudes.
About 16\% (153) sources are not in the JVAS sample and have been
observed with the VLA in B configuration at 8.4 GHz as part
of the CLASS survey, in a similar manner as the targets in the JVAS sample
(Myers et al. 2001). These provided us with radio positions at
an accuracy similar to those of the JVAS sources.
Fifty-eight sources are fitted by multiple components at 8.4 GHz.
For these, the positions of the brightest components are used.
Fifteen objects are not detected with the VLA, all exhibiting extended
structure on arcminute scales.

\subsection{The completeness and reliability of JVAS}

JVAS contains 762 sources which overlap with JVAS++. 
In the same area of sky, another 371 sources are part
of JVAS, but are not in JVAS++.
In addition, 153 objects are in JVAS++, but are
not part of JVAS.

The large differences between the JVAS and JVAS++ 
are due to several effects:
\begin{itemize}
\item[1] the use of different selection surveys for JVAS.
\item[2] the use of a difference flux density scale for JVAS, which
was found to be offset by 10\% at 5 GHz.
\item[3] the use of a variety of spectral data in the target
 selection, the non-detection of some sources, and the exclusion of some
sources for which the observed VLA position was found to be offsetted by 
more than 40 arcseconds from the pointing position.
\end{itemize}

\begin{figure*}
\hbox{
\psfig{figure=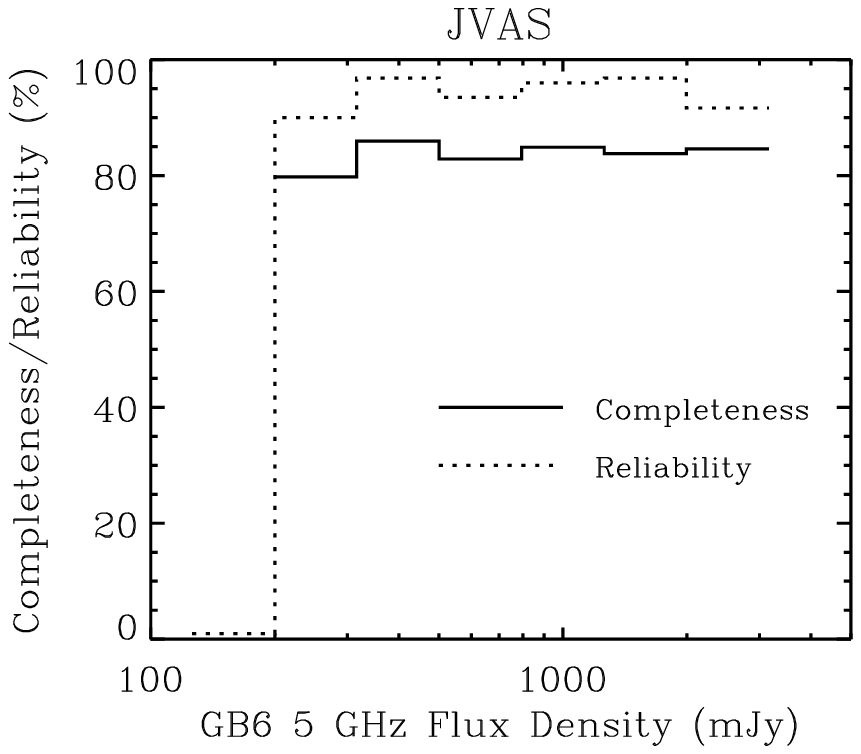,width=8cm}
\psfig{figure=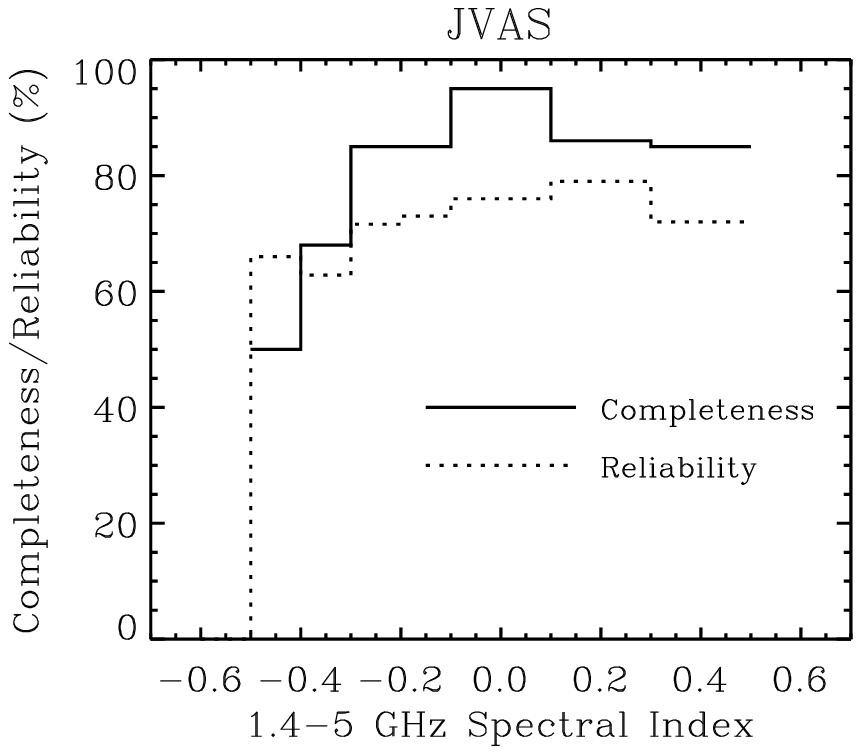,width=8cm}}
\caption{The completeness and reliability of the JVAS survey as 
function of GB6 5 GHz flux density (left) and 1.4-5 GHz Spectral index (right),
with respect to JVAS++ as defined in section 2.\label{relcomp}}
\end{figure*}
\begin{figure*}
\psfig{figure=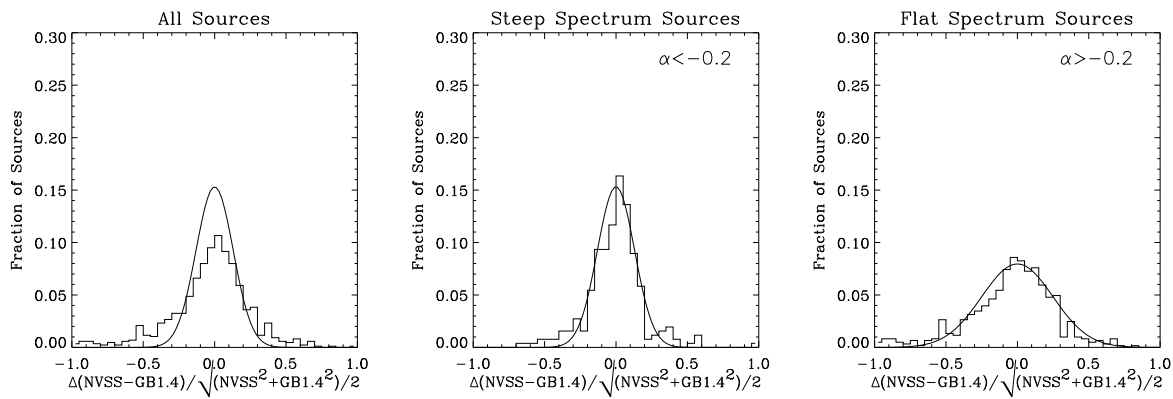,width=16cm}
\caption{\label{nvssgb1400} The distribution of the normalised differences in 
1.4 GHz flux density of the GB1400 and NVSS catalogues for all the sources in 
the complete sample, (left), steep spectrum sources (middle) and flat 
spectrum sources (right).}
\end{figure*}
\begin{figure*}
\psfig{figure=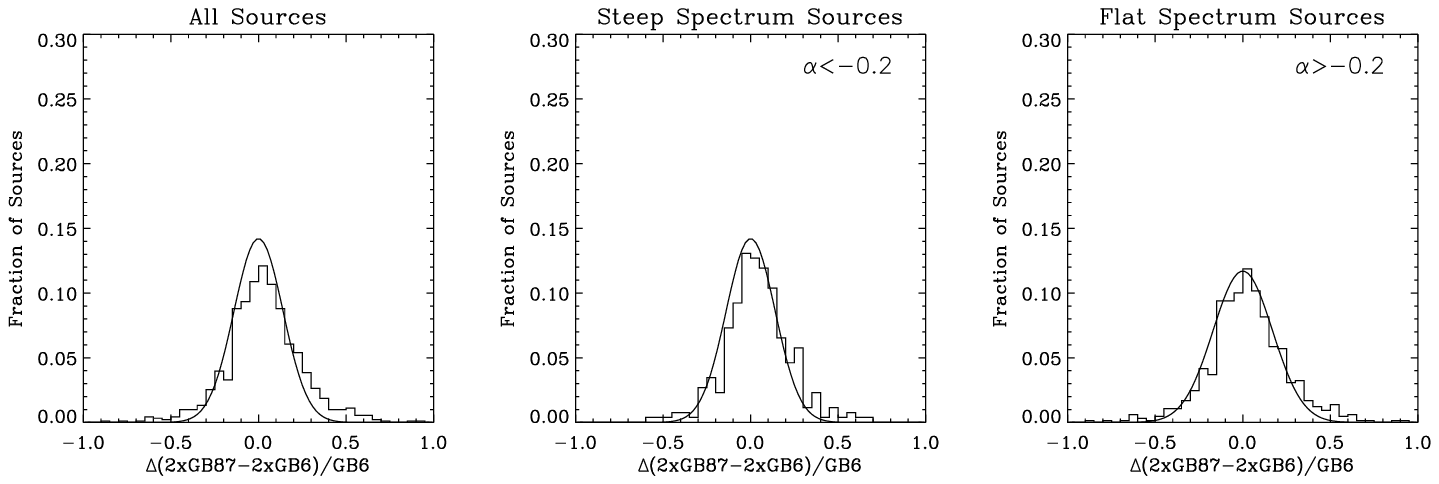,width=16cm}
\caption{\label{gb6cbs} The distribution of the normalised differences 
in 5 GHz flux density of the GB6 and 87GB catalogs for all the sources in 
JVAS++,
(left), only the steep spectrum sources (middle), and 
only the flat spectrum sources (right)}
\end{figure*}

 The use of a different flux density scale at 5 GHz, implies
that the actual selection criteria for JVAS are, $S_{5 GHz}>180$ mJy,
and $\alpha_{1.4-5}>-0.58$. This effect, in combination with the inclusion
of sources with 150 mJy $<S_{5GHz}<$ 180 mJy, to fill holes in the 
sky distribution, causes the low reliability of JVAS.
Note that since these sources are fainter than 200 mJy, they do not influence
the reliability above this flux density, which is 
typically $>95\%$ (see fig. \ref{relcomp}, left). However, they 
are included, when the reliability is plotted as function of spectral index
(fig. \ref{relcomp}, right), causing it to be typically $<75\%$.

The use of different selection surveys has three effects.
Firstly, the Greenbank 1400 MHz survey, used for the selection 
of JVAS, has a 12$'$ FWHM resolution, compared to 45$''$ for the 
NVSS and a search radius of 70$''$ used for the revised `complete' sample. 
Therefore, sources with extended structure at arcminute scales
may have been excluded from JVAS with $\alpha<-0.5$, but included 
in the comparison sample. This is reflected in Figure \ref{relcomp}
(left), showing that the completeness of JVAS is decreasing
to $\sim 50\%$ at the steep spectral index cut-off.
Indeed 49 objects exhibit extended structure at arcminute scales.
These are shown in the appendix, where their NVSS images
are overlaid with data from the Digitized Sky Survey (Lasker et al. 1990).

Secondly, the use of different selection surveys may cause
sources to be included or excluded due to their possible 
intrinsic variability. Thirdly, measurement uncertainties may
cause differences in the selection. 
To show the influences of variability and measurement 
uncertainties on the selection process, the Greenbank 1400 MHz
observations were compared with the NVSS, and the 87GB data were compared
with the GB6.
Figure \ref{nvssgb1400} shows the normalised differences between the GB1400 
and NVSS flux densities for the sources in the complete sample, with
no radio spectral selection on the left, `steep' spectrum ($\alpha <-0.2$) 
sources in the middle, and `flat' spectrum ($\alpha >-0.2$) sources in
the right panel. 
About 12\% of the sources do not have a GB1400 flux density since 
they are too faint to be in the catalog by White \& Becker (1992). These
are excluded from the analysis.
The NVSS flux densities were decreased by
3\% to match the GB1400 flux density scale. The distributions are overlaid by
Gaussians with $\sigma=$ 0.13, 0.13, and 0.25 respectively.
The distribution for the flat spectrum sources is clearly much broader ($\sigma=0.25$) than that for steep spectrum objects ($\sigma \sim$ 0.13).
This indicates that the differences between GB1400 and the NVSS for flat 
spectrum sources is dominated by variability, which is about 2 times larger
than the measurement uncertainties.
From a similar distribution of normalised flux density differences between
NVSS and the FIRST survey (White et al. 1997), we found that the 
measurements errors in both FIRST and NVSS are about 5\%. 
This means that, assuming that the distribution of the 
steep spectrum sources is solely due to measurement errors,  
the GB1400 survey has an uncertainty in flux density of 12\%. 
Note that this uncertainty has a proportional term and a constant
term, the latter due to noise and confusion. In FIRST and NVSS, this
constant term is unimportant for our sample. However for the 
GB1400 survey with
a noise level of $\sim25$ mJy,  this factor probably dominates  
the uncertainty in flux density for the faintest sources.  

A similar analysis has been carried out for the two selection surveys at
5 GHz, which is shown in figure \ref{gb6cbs}. The GB6 and 87GB are not
independent; GB6 is based half of on the 87GB dataset, and half on 
a similar dataset, taken a year earlier. Since we are interested in 
the variability aspect, we assumed that the flux densities of 
the sources in the second dataset is $2\times GB6-87GB$.
The normalised differences between the flux densities taken
at these epochs are shown in figure \ref{gb6cbs}, with 
on the left with no spectral selection, in the middle for steep
spectrum sources, and at the right for flat spectrum sources.
The distributions are overlaid with a Gaussian with a $\sigma$ of 
0.14, 0.14, and 0.17 respectively. Assuming that the distribution of the 
steep spectrum sources is solely due to measurement uncertainties, each 
dataset has a typical error of 10\%, and the uncertainty in flux
density of the GB6 is 7\%. The constant noise term of GB6 is 
expected to contribute for only 2\% for the faintest sources in the
sample. 
Note that the flat spectrum sources seem to be only slightly variable, 
compared to the data at 1.4 GHz, although flat spectrum sources
are known to be increasingly variable towards higher frequencies.
This is due to the fact that the time-baseline at 5 GHz ($\sim$ 1 year) is much
shorter than at 1.4 GHz ($\sim$ 10 years).

It is difficult to estimate the completeness and reliability of JVAS
and JVAS++, due to all the effects described above. 
Most of the differences between the two samples are not caused by 
incompleteness or unreliability of JVAS, but due to a difference in 
selection criteria. To assess the completeness of 
JVAS and JVAS++, we performed a simple simulation.
We treated the NVSS and GB6 data as infinitely accurate and 
added Gaussian distributed offsets (with $\sigma$'s as estimated above)
to the flux densities, and determined how many sources enter and leave the 
sample. In this way, the completeness and reliability of JVAS++
were estimated both to be 95\% respectively. For JVAS, we
changed the selection criteria to $\alpha_{1.4-5GHz}>-0.58$ and $S_{5GHz}>180$
 mJy after the offsets were added, to mimic the use of a different 
flux density scale. In addition it was assumed that all the sources 
with $S<160$ mJy were added artificially in JVAS to fill in holes 
in the sky distribution ($\sim$100 objects). In this way the 
completeness and reliability of JVAS was estimated to be 90\% and 
70\% respectively.\footnote{Note that all the sources in JVAS are real and 
that their positions are reliable}

\section{The automated optical identification procedure}
\subsection{The APM $-$ POSS-I catalogue}

Optical identification of sources in the JVAS and JVAS++ samples 
was carried 
out using the output
catalogue of the APM scans of the Palomar Sky Survey (POSS-I) photographic
plates in the $e$ (red) and $o$ (blue) passbands (Bunclark \& Irwin, 1983). 
Plates were scanned with $\sim 0.5''$ pixels. 
The APM covers the Northern sky with $|b|\ge20^\circ-30^\circ$.
For each detected object a magnitude was measured and the image classified
for the $e$ and $o$ plates separately. The images were classified as galaxies,
stellar, merged objects or noise based on combinations of various 
image parameters such as moments, ellipticities etc. 
The spread in the distribution of each parameter with respect to the stellar 
locus was used to weight that parameter when combined to form
a final classification parameter. The distribution of the combined 
parameter has a mean and spread which is dependent on magnitude, and a 
statistical correction was calculated to convert these for the stellar objects
 to Gaussian distributions with zero mean and rms=1. This correction was then
applied to all images and the standard deviation from a stellar profile 
$N\sigma_c$ was recorded. For images about 2 magnitude above the plate limit, 
this classification is $\sim90\%$ accurate.

The $e$ and $o$ catalogues were merged, assuming that sources within 2$''$ 
are one and the same object. Objects with centroids on the $e$ and $o$ plates which
differ in position between 2 and 4$''$ were considered to be ``nearly'' 
matches.
In such case only the position on the E plate were preserved.
The final positions of the objects were corrected for subtle distortions
of the plates which are most prominent near the plate edges. 

The zero-point for photometry in each field was calculated by
assuming that the $e$ plate has a limiting magnitude of 20.0 and by assuming a 
universal, magnitude independent position in the colour-magnitude plane
for the stellar locus, as described by McMahon \& Irwin 1992. 
The zero-point of the magnitude system has an rms accuracy of 0.3 magnitude.

The APM $-$ POSS-I catalogue can be accessed via the internet at 
{\it www.ast.cam.ac.uk/$\sim$apmcat}.

\subsection{The optical-radio correlation}

\begin{figure*}
\hbox{

\psfig{figure=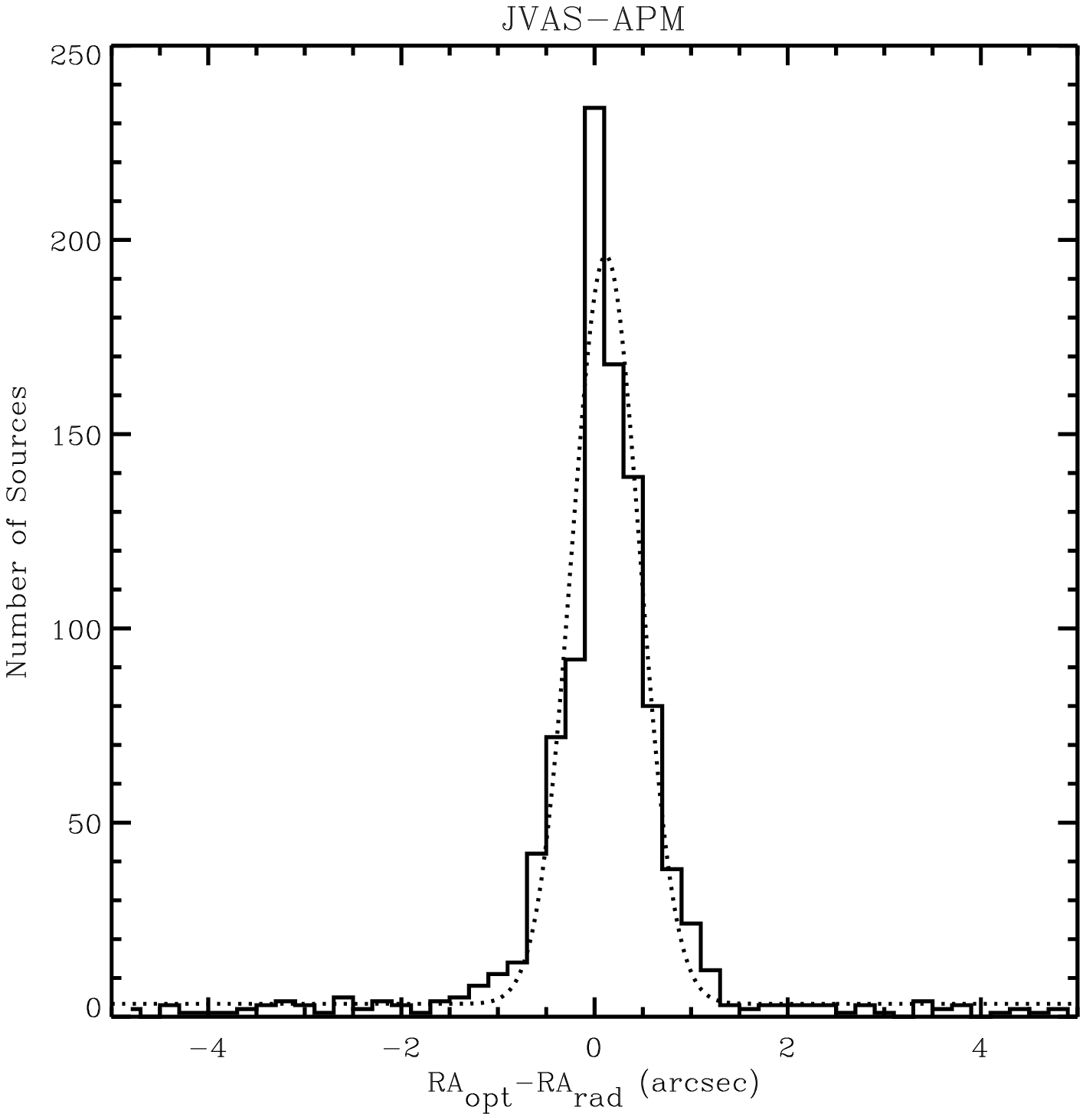,width=8cm}
\psfig{figure=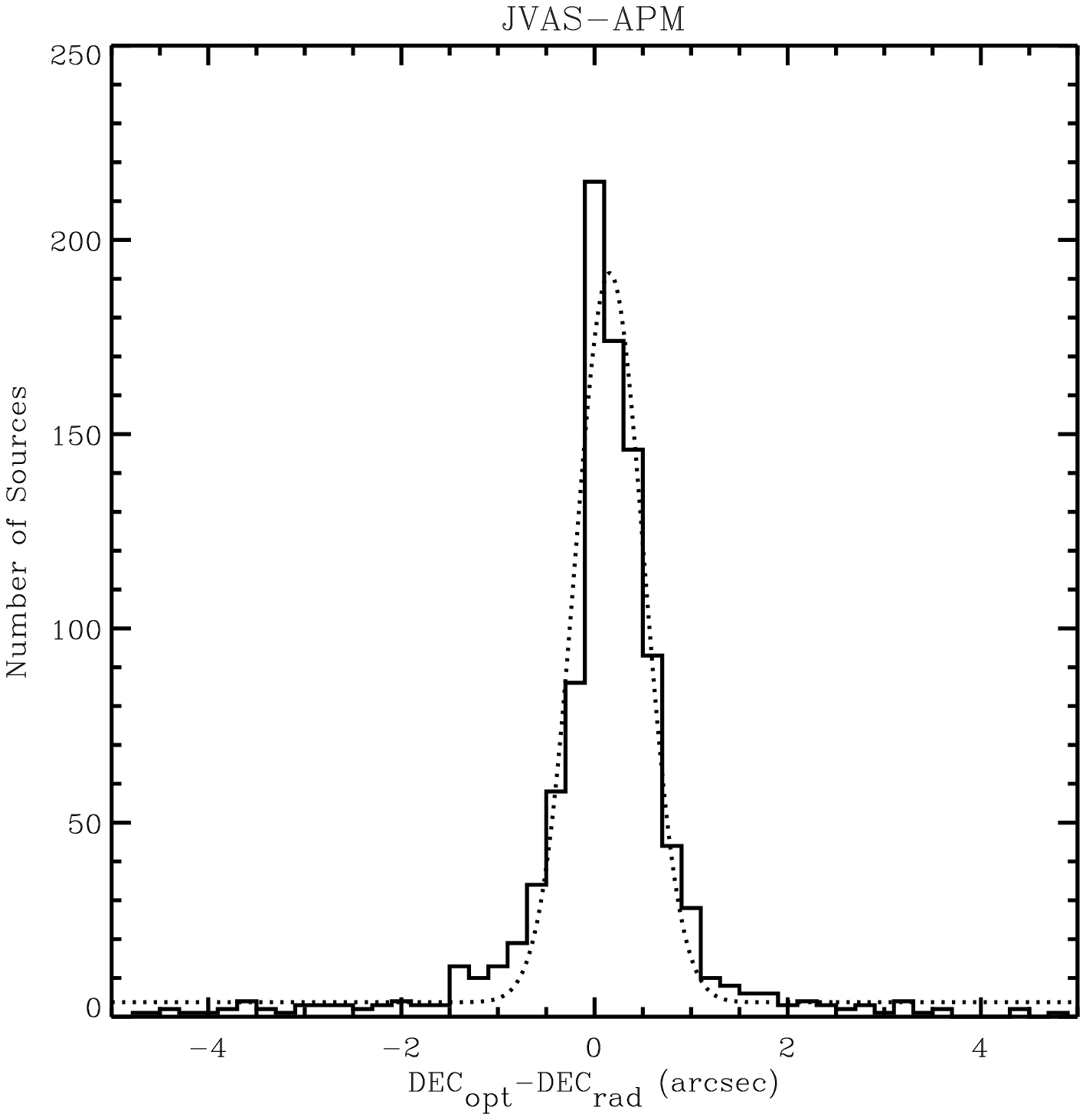,width=8cm}
}
\caption{\label{positions}The distribution of 
$\Delta \alpha_{opt} - \alpha_{rad}$ and $\Delta \delta_{opt} - \delta_{rad}$,
for all the optical objects to a radio position of a JVAS++ source.}
\end{figure*}

Three source lists were constructed:
\begin{itemize}
\item[I:] Sources which are part of both  JVAS++ and the JVAS sample
(762 objects)
\item[II:] Sources which are only part of JVAS++ (153 objects)
\item[III:] Sources which are only part of JVAS (371 objects)
\end{itemize}
Hence, the combination of sources in lists I and II make up JVAS++, 
and the sources in lists I and III make up the JVAS sample 
(in the selected area of sky). 
The three lists were correlated with the APM$-$POSS-I catalogue, selecting
all optical objects within $30''$ of each 8.4 GHz radio position. 
For the objects with
extended radio structure at arcminute scales, the radio-optical overlays, 
as shown in the appendix, were used to search for a possible 
 bright optical identification. These were added manually.
The distributions of 
$\Delta \alpha_{opt-rad}$ and $\Delta \delta_{opt-rad}$ are shown in figure \ref{positions}, for sources in the JVAS++ sample,
excluding objects which exhibit multiple components at 8.4 GHz and/or 
extended structure at arcminute scales.
From these plots the accuracy of the APM positions was determined, by
fitting the $\Delta \alpha$ and $\Delta \delta$ distributions with a 
Gaussian on top of a flat distribution, as expected for a combined 
population of genuine identifications and random background objects.
It shows that the APM positions have an uncertainty of rms=0.4$''$
in both $\alpha$ and $\delta$, and show no systematic offsets to the radio
positions. 

\begin{figure}
\psfig{figure=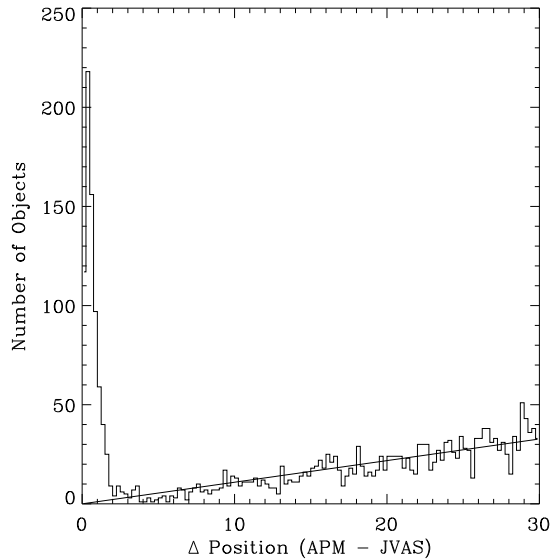,width=8cm}
\caption{\label{rpos} The offsets between the JVAS++ positions and the optical
APM positions. The solid line indicates a fit to the background source 
density of 0.0006 arcsec$^{-2}$.}
\end{figure}

Figure \ref{rpos} shows the distribution of sources around the JVAS++ 
positions. The solid line is a fit to the background source density
of 0.0006 arcsec$^{-2}$. We found that this background level was
not enhanced by possibly related cluster objects around the radio sources,
by counting the number of objects within a 30$''$ radius, offset 5$'$ north
from each source. 
It is therefore expected that $\sim 2\%$ of the sources have 
a random background object within 3 arcsec from their radio position.
Such a background object will be incorrectly identified as the 
optical identification unless the genuine identification is detected and 
has an APM position closer to the radio position. 

Unfortunately, for several reasons the errors in the optical positions are 
not Gaussian distributed, as assumed above:
\begin{itemize}
\item[1] {\bf Bright galaxy identifications:} The central positions of bright
extended objects, e.g. nearby galaxies, can be determined 
significantly less accurate in the APM than those of  fainter stellar objects.
\item[2] {\bf Merged Objects:} Due to the limited resolution of the APM
scans, two or more nearby sources are sometimes merged into one single 
object. In such a case, the central position of the merged object can be 
several arcseconds away from the position of one of the individual 
objects (i.e. the identification).
\item[3] {\bf Bright Stars:} Bright Stars, can obliterate 
other objects in their vicinity, or can produce spurious images.
\item[4] {\bf Anomalies in the APM Catalogue:}. 
In very rare cases the APM can produce 
nonsensical results unrelated to any of the former effects, e.g. plate defects
such as dirt and scratches, aeroplane tracks and asteroids.
\end{itemize}

These effects produce a broad non-Gaussian tail in the error distribution
at a $\sim 5-10\%$ level. To identify these problems for the individual 
objects, all the APM images were checked by eye and compared 
with images from the Digital Sky Survey as retrieved from Skyview.
In the course of this process the images were also searched for possible
surrounding groups or clusters of galaxies. In addition, if no identification
in the APM was found, it was checked whether a possibly faint  
identification was visible in the DSS. 
All the JVAS and JVAS++ sources which were found to have one of these problems 
are shown in the appendix. 

\subsection{Completeness and reliability of the optical identifications}

\begin{figure}
\psfig{figure=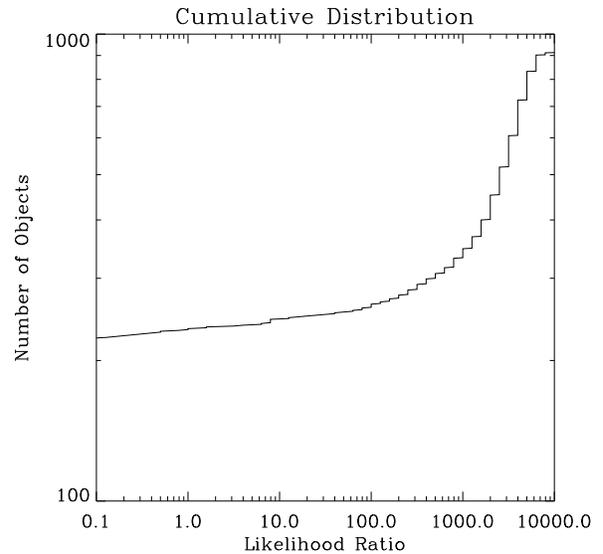,width=8cm}
\caption{ \label{LHR} The cumulative distribution of likelihood ratios 
for the complete sample.} 
\end{figure}

The likelihood method 
(de Ruiter, Arp and Willis, 1977) was used to quantify the 
identification procedure.  First, for each candidate identification
the following dimensionless measure of the uncertainty in position
difference, $r$, was calculated:
\begin{equation}
r = \sqrt{\frac{\Delta \alpha ^2}{\sigma _{\alpha opt}^2} 
+ \frac{\Delta \delta ^2}{\sigma _{\delta opt} ^2}}
\end{equation}
where $\Delta \alpha$ and $\Delta \delta$ are the offsets between the 
optical and radio positions, and $\sigma _{\alpha opt}$ and 
$\sigma _{\delta opt}$ the uncertainties in the 
optical right ascension and declination positions. The error in the radio 
position is small $<0.05''$ compared to the optical error ($0.55''$) 
and therefore neglected. 
Given the normalised position difference $r$ for a certain 
radio-optical pair, the likelihood ratio is defined by 
(de Ruiter, Arp and Willis, 1977)
\begin{equation}
LR(r) = \frac{dp(r|id)}{dp(r|c)} = \frac{r \  e^{\frac{-r^2}{2}}}{2 \lambda \
  r \ e^{-\lambda r^2}} = \frac{1}{2\lambda} e^{\frac{r^2(2\lambda-1)}{2}}
\end{equation}
where $\lambda = \pi \sigma_{\alpha opt} \sigma_{\delta opt} \rho_{bg} =
0.5 \rho_{bg}$, where $\rho_{bg}$ is the background source density 
which is stored for each POSS-plate in APM calibration files.  
This is the ratio of the probability that a given object
found between $r$ and $r+dr$ is the correct identification $p(r | id)$, divided
by the probability that it is a contaminating object $p(r | c)$.
The cumulative distribution of likelihood ratios is shown in figure \ref{LHR}.

The likelihood ratio cutoff used for this sample is 1.0
which means the probability that the given object
is the correct identification is at least equal to the probability 
that it is a background object.
Using this cutoff we find an identification rate, $\theta$, of 83\%
(i.e. 756 out of 915). 
This identification fraction can be related to the two 
{\it a posteriori} probabilities that the object found at an angular 
distance $r$ from
the radio source position is a genuine identification, $p(id|r)$, or a 
confusing object, $p(c|r)$, using Bayes' theorem

\begin{equation}
p(id|r)=\frac{\theta \ LR(r)}{\theta \ LR(r) + 1 - \theta}   \ \ \ \ \ 
p(c|r)=\frac{1}{\theta \ LR(r) + 1 - \theta}
\end{equation}

It was assumed that the sources which were found to be
merged objects or bright galaxies have an infinite likelihood ratio, 
resulting in $p(id|r)=1$ and $p(c|r)=0$. 
The likelihood ratios of the objects, for which the manual check 
has shown they have a genuine optical identification but a large
radio-optical position offset, are set to $2\times10^3$, as if the offset
was zero, and therefore as
100\% reliable identifications.
The completeness, $C$, of the identifications, which is the number of 
accepted identifications over the number of correct id's, 
and the reliability, $R$, which is the fraction of identified sources
for which the identification is correct, are given by:
\begin{equation}
C=1-\biggl( \sum_{L<1} p(id|r)  \biggr) / N_{id} 
\end{equation}
\begin{equation}
R=1-\biggl( \sum_{L>1} p(c|r)  \biggr) / N_{id}
\end{equation}
where, $N_{id}$ is the total number of identifications.
These correspond to a completeness $C=99\%$ and a reliability $R=99\%$
for LR$\ge1$ used.

\section{Results and discussion}

Three tables were produced for the three source lists, which are
shown in the appendix, with a detailed description of 
the columns. 

\subsection{Radio spectral properties and optical identifications}

\begin{figure}
\psfig{figure=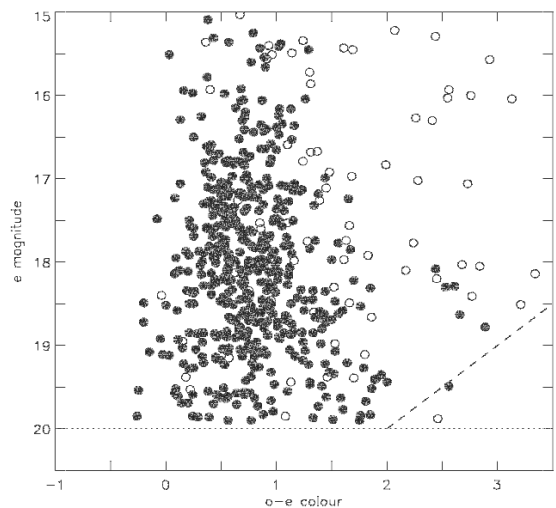,width=8cm}
\caption{\label{magcolor} The red $e$ magnitude as function of $o-e$ colour
for extended (open circles) and stellar (filled circles) 
identifications. The dotted 
and dashed lines indicate the limit in $e$ magnitude and the average limit
in $o$ magnitude (which differs per plate) respectively. }
\end{figure}

\begin{figure}
\psfig{figure=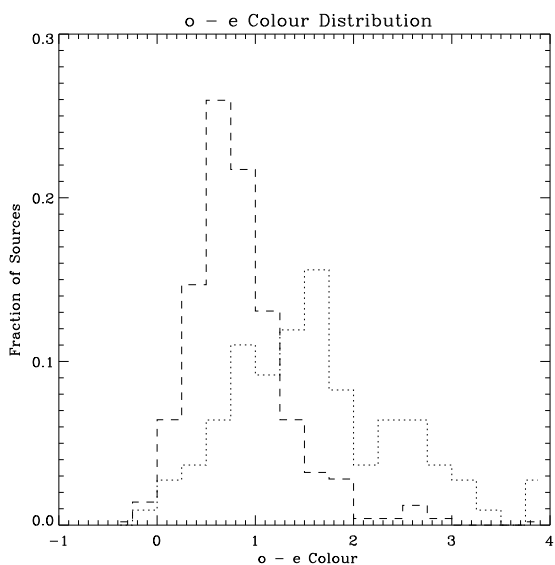,width=8cm}
\caption{\label{colordis} The normalised $o-e$ colour distribution for the 
JVAS++
identifications. The dashed and dotted lines represent the 
stellar and the extended objects respectively.}
\end{figure}

The JVAS++ sample is uniquely suited to study the high luminosity end
of the flat spectrum radio source population. 
Figure \ref{magcolor} shows the $e$-magnitudes of the optical identifications
as function of their $o-e$ colours, classified as stellar (filled circles) and 
extended (open circles) objects on the red plates. Objects with
only limits on the colours are not shown.
The colours of the extended 
objects are more biased towards the red than those of the stellar 
identifications, as expected for quasars and galaxies, 
which are found to have generally blue and red optical colours respectively, 
This can also be seen in figure \ref{colordis} where the
colour-distribution of the two classes of objects are shown.
Only at $e>19$, near the plate limit, does this picture get blurred, since
no reliable classification can be made and the large majority of objects are 
classified as stellar.

\begin{figure}
\psfig{figure=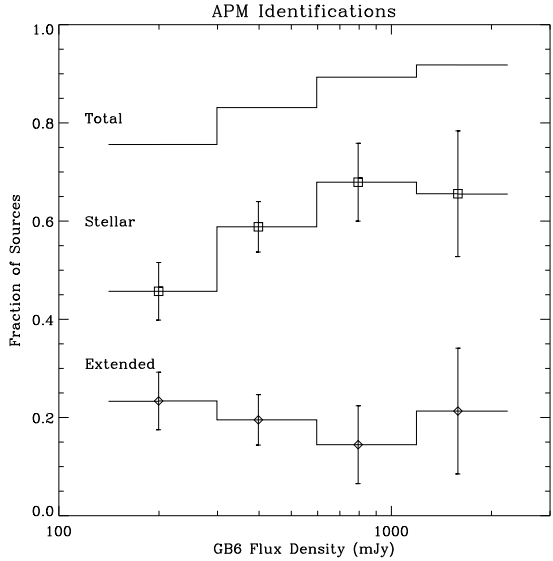,width=8cm}
\caption{\label{idfrac} The fraction of JVAS++ sources identified with
APM objects (total), APM objects classified as stellar and 
classified as extended in the red band, as function of GB6 flux density.
The uncertanties for the total sample are similar to those for
the stellar objects.}
\end{figure}

Figure \ref{idfrac} shows the optical identification fraction as function of 
GB6 flux density. The upper line indicates the total identification fraction
(including extended, stellar and merged objects). The middle line indicates
the fraction of JVAS++ sources identified with APM objects classified as
stellar in the red band.  The lower line indicates
the fraction of JVAS sources identified with APM objects classified as 
extended in the red band. 
There is a  hint that both the total and stellar identification
fraction decrease with decreasing flux density.
These two effects are likely caused by a change-over to galaxy 
identifications at faint flux density levels, which leads to 
more loss against the plate limit. A similar effect at a similar 
flux density level has been seen by Shaver et al. (1997) and
Falco, Kochanek \& Mu\~noz (1998).
Note that due to the uncertain classification near the plate limit, 
the true decrease in identification fraction for quasars with flux density 
may even be more pronounced.

\begin{figure}
\psfig{figure=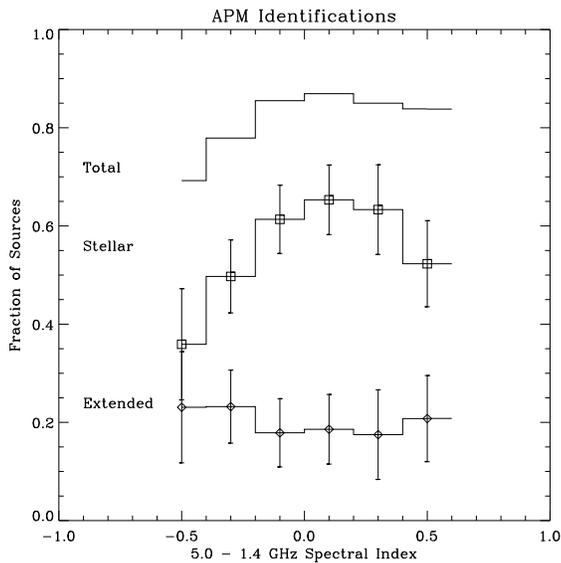,width=8cm}
\caption{\label{sifrac}The fraction of JVAS++ sources identified with
APM objects (total), APM objects classified as stellar and 
classified as extended in the red band, as function of 8.4-1.4 GHz spectral
index.}
\end{figure}

Figure \ref{sifrac} shows the optical identification fraction as function 
of 5.0 to 1.4 GHz spectral index. The lines indicate the specific 
identification fractions
as in figure \ref{idfrac}. Clearly, the total and stellar identification
fractions decrease towards steeper spectral indices, while the
fraction of sources identified with extended objects appears to be
independent of spectral index. This trend can be explained by
assuming that at $\alpha<-0.3$, the unbeamed population of `steep' spectrum 
galaxies contributes significantly to the total radio source population. 
Most of these radio galaxies will be too faint to appear on the 
POSS, and will be unidentified radio sources. 
This results in a drop in the total identification fraction and the 
identification fraction of quasars.

\begin{figure}
\psfig{figure=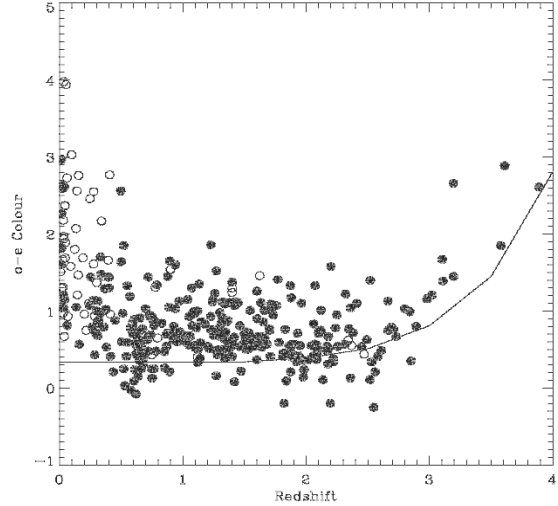,width=8cm}
\caption{\label{zcolor} The $o-e$ colour of extended (open circles) and stellar
(filled circles) objects as function of redshift, for all sources with an available
redshift.The solid line indicates the expected $o-e$ colour of an average 
quasar reddened by intergalactic absorption (see text). }
\end{figure}

For the optical identifications with available redshifts in the literature,
the $o-e$ colours are shown as function of redshift (figure \ref{zcolor}).
This figure should be interpreted with great care, since it includes 
a strong observational bias, e.g. towards bright galaxies and red
quasars. Therefore, not surprisingly, most of the extended optical
objects (galaxies) are observed to be at $z<0.5$. They are
clearly redder than the stellar objects at similar redshifts.
Note however that a few objects which are classified as extended, 
are found at much larger redshifts.
Their optical spectra show that they are actually quasars, 
wrongly classified as extended objects in the APM.
There is a clear trend that the optical colours of the quasars 
become redder towards high redshifts. This is a well known effect and
the forms the basis of the colour selection of candidate high
redshift quasars (Hook et al. 1996), and is due to 
intervening Ly$\alpha$ absorption systems.
Note however, that the majority of the objects in this 
redshift regime was actually selected for spectroscopic follow up
on the basis of their red colour ($o-e\ge 1.0$), 
which may have strengthen this effect.
The solid line indicates the expected $o-e$ colour of an average quasar 
reddened by intergalactic absorption. As a quasar spectrum, a power
law with spectral index of $-$0.5 was used with emission lines taken
from the composite quasar spectrum as constructed by Francis et al. (1991).
For the intergalactic absorption the model of Madau (1995) was used.
The model follows the trend of redder colours towards high redshift well.
At low redshift the data are systematically redder than the model.
This is probably due to a contribution of underlying galaxy light.

\begin{figure*}
\hbox{
\psfig{figure=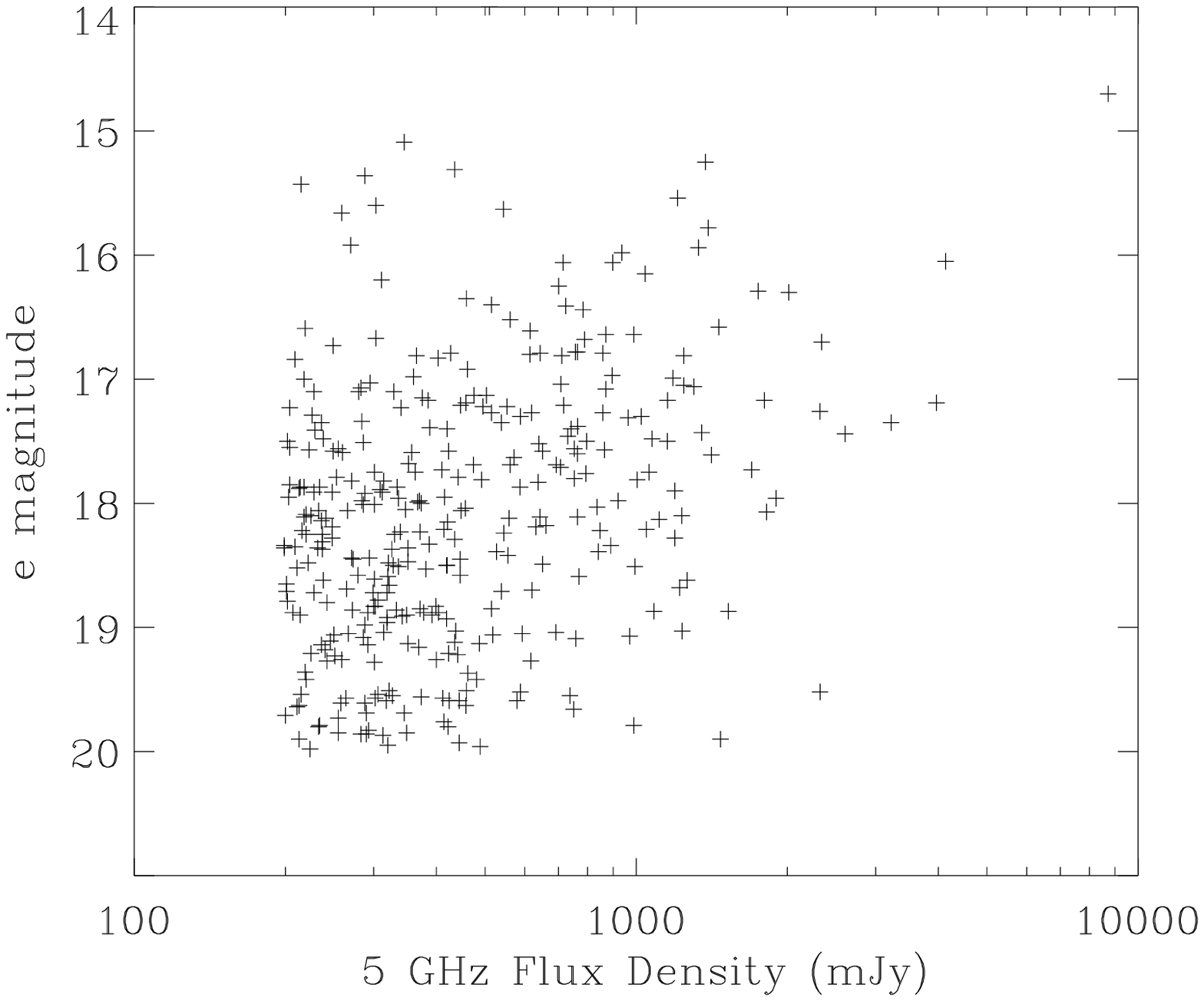,width=9cm}
\psfig{figure=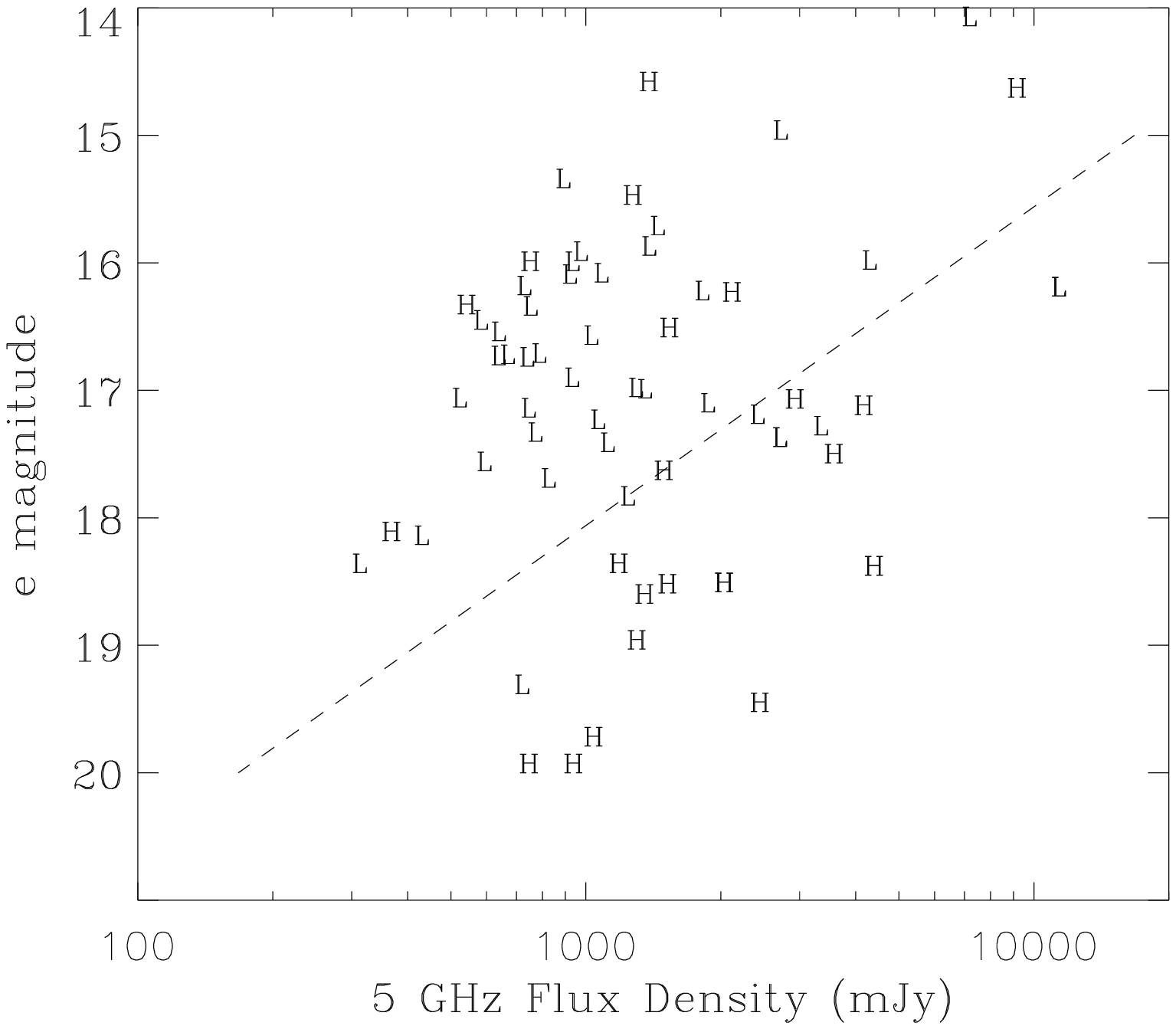,width=9cm}
}
\caption{\label{mag_radio} (left) The $e$ magnitude versus 
of 5.0 GHz flux density for radio sources 
identified with blue $o-e<1.0$ stellar objects. (right) The $e$
magnitude versus of 5.0 GHz flux density for radio sources 
in the sample, which have optical polarisation measurements in the 
literature, with 'L' and 'H' denoting low and high polarisation
quasars. Objects at $z<0.5$ have been omitted. The dotted line 
indicates an optical-to-radio spectral index of $-$0.75.
}
\end{figure*}

\begin{figure}
\psfig{figure=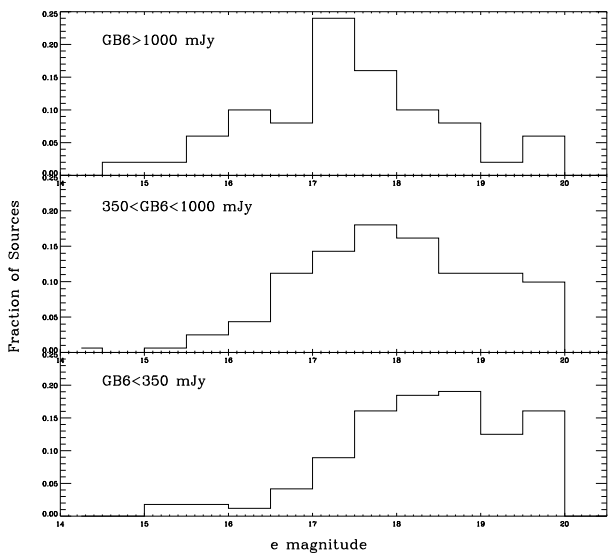,width=8cm}
\caption{\label{optdistr} The magnitude distributions for bright 
(GB6$>$1 Jy), intermediate (350 mJy $>$ GB6 $>$1 Jy), 
and faint (GB6 $<$ 350 mJy) radio sources, optically identified with
blue $o-e<1.0$ stellar objects.}
\end{figure}

It is interesting to investigate the relation 
between the optical apparent magnitude of a quasar and 
its radio flux density, since both should be related to the 
power output of the central engine. Bolton \& Wall (1970) showed
that objects at fainter flux densities tend to have fainter 
optical magnitudes, but that there is a large spread. 
This trend is also present in figure \ref{mag_radio} (left), 
where the 
e-magnitude is shown as function of 5 GHz flux density for sources 
which are classified as stellar on the red plates, and which have
a blue $o-e<1.0$ colour. 
This can also be seen in figure \ref{optdistr}, 
where for the same subsample, the magnitude distributions
are shown for bright (GB6$>$1 Jy), intermediate (350
mJy $>$ GB6 $>$1 Jy), and faint (GB6 $<$ 350 mJy) radio sources.
 
It is interesting to investigate what may cause this  
large spread in the ratio of optical to radio luminosity.
We noticed that many of the optically faintest quasars were
known to be High Polarised Quasars (HPQ); 
objects which exhibit optical polarisation at levels $>3\%$.
In the right panel of figure \ref{mag_radio} the $e$ magnitude
versus 5 GHz flux density is shown for objects in the sample
which have their optical polarisation properties studied 
by Fugmann \& Meisenheimer (1988) or Wills et al. (1992).
Indeed, it shows that the HPQ ('H') prefer to have high 
radio to optical flux ratios while the low polarised
quasars (LPQ; 'L') show low ratios. 
Objects at $z<0.5$ were excluded from this plot, to avoid
 possible host galaxy contamination, and to exclude
the population of nearby BL Lac objects and optically
violent variable (OVV) quasars which also
show high optical polarisation, but often combined with 
low or rapidly varying radio to optical flux density ratios.

The distribution of HPQs and LPQs is consistent with the 
idea that the optical quasar light is a combination of 
a polarised synchrotron component plus an unpolarised component
(e.g. Smith et al. 1994). 
If the unpolarised component is significantly brighter than
the synchrotron component (LPQ), 
then the radio flux (also synchrotron) 
will be relatively faint compared to the optical, resulting in a 
`flat' optical to radio spectral index. If the unpolarised component
is fainter (HPQ), and a constant spectral index is assumed
for the synchrotron component, then the radio flux will be 
relatively bright, resulting in an overall `steep' radio-to-optical 
spectral index.
Indeed, the dashed line in figure \ref{mag_radio} (right) indicates
an optical-to-radio spectral index of $-$0.75, with most of 
the LPQs and HPQs situated above and below this line respectively. 

\begin{figure}
\psfig{figure=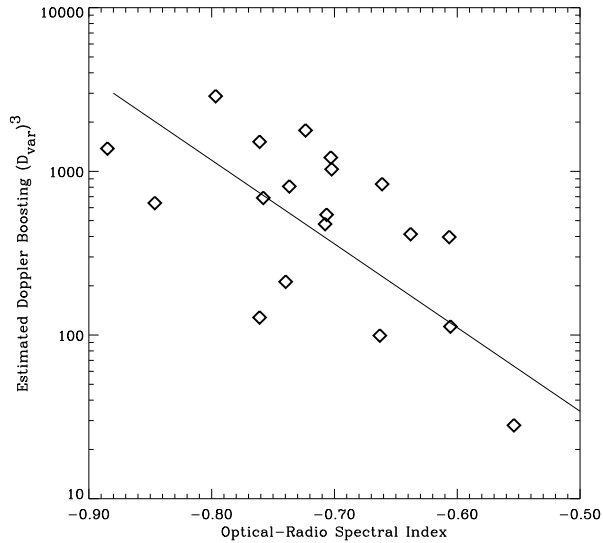,width=8cm}
\caption{\label{boosting} The Doppler boosting, 
estimated by L\"ahteenm\"aki \& Valtaoja (1999) 
as function of radio to optical spectral index, for bright 
JVAS sources. The line indicates the expected relation if the 
 change in spectral index is due to Doppler boosting of only 
the radio emission, leaving the optical unaffected. }
\end{figure}

The most straightforward explanation of
why the ratio of the polarised to unpolarised components 
varies so much from quasar to quasar is Doppler boosting 
of the synchrotron component (eg. Wills et al. 1992).
We therefore hypothesise that the large spread in optical to 
radio luminosity ratios is caused by source to source 
variations of Doppler boosting of the radio flux, leaving the 
unpolarised component of the optical emission unaffected.
The fact that in our sample we see the HPQs fainter than
the LPQs is counter-intuitive since these are the boosted
objects. However, in this scheme, the HPQs are not the boosted 
counterparts
of the LPQs in the sample, but are the boosted counterparts
of a population which is fainter in the optical and radio. 
 If this explanation is correct, then a correlation is expected
between the Doppler factor and the optical to radio spectral index.
L\"ahteenm\"aki \& Valtaoja (1999) estimated the Doppler factors of 
eighty of the brightest flat spectrum objects in the sky, using 
total flux density variation monitoring data at 22 and 37 GHz.
Thirty-nine objects in their sample overlap with JVAS.
Nineteen of those are located at
$z>0.5$, and  have APM identifications with $o-e<1.0$ (to avoid possible influence of the host galaxy and extinction).
For the optical and radio 
flux densities, the average of $o$ and $e$ magnitudes, and the 
average of 1.4, 5.0 and 8.4 GHz flux densities are used, to 
minimise the influence of variability. 
The estimated Doppler boosting, $D_{var}^3$, is plotted against the 
optical-to-radio spectral index in figure \ref{boosting}.
It shows that the Doppler factor is indeed 
correlated (97\% significance) with the radio-to-optical
spectral index, and that the slope of the relation is as 
expected as for the hypothesis that 
the large spread in optical to 
radio luminosity ratios is caused by source to source variations of 
Doppler boosting of the radio flux leaving most of the 
optical emission unaffected.
Evidently, this scheme is too simplistic. First of all, 
the spectral index of the synchrotron component is most likely
steeper in the optical than in the radio, causing the boosting 
to be stronger in the optical than in the 
radio, changing the observed optical-to-radio spectral index 
of the synchrotron component. Furthermore, in the optical we may see 
a related, but younger part of the synchrotron component,
which possibly exhibits different boosting properties, complicating the 
simple picture scetched above.

However, the main idea is supported by two previous studies. 
Firstly, Yee \& Oke (1978) and Shuder (1981) 
showed that the emission line luminosity for 
a range of AGN types is proportional to the luminosity of the 
underlying optical continuum over four orders of magnitude. Secondly, 
Rawlings \& Saunders (1991) found that the emission line luminosity of an 
unbiased sample of FRII radio galaxies is approximately proportional to 
the total jet kinetic power, which is closely coupled to the power of 
the central engine. This implies that the optical luminosity should also
be a direct indicator of the jet power, hardly affected by Doppler 
boosting. 
Indeed, Wills \& Brotherton (1995) use the ratio of the radio core to 
optical continuum luminosity, $R_V$, 
which is the equivalent to the radio-optical
spectral index as used in this paper, as an improved measure of 
quasar orientation over the ratio of radio-core to lobe flux density, $R$. 
They show that the use of $R_V$, rather than $R$, results in a significantly 
improved inverse correlations with the beaming angle as deduced from 
apparent superluminal velocities and inverse-Compton-scattered X-ray emission,
and with the FWHM of a quasar's broad H$\beta$ emission line.
The use of optical continuum luminosity rather than extended radio luminosity 
to represent the unbeamed jet power probably works better because the latter
is more affected by source-to-source differences in the intergalactic medium
(Wills \& Brotherton, 1995), and source age.

\section{Summary}

We have described the automated optical identification procedure of 
the sources from the Jodrell Bank $-$ VLA Astrometric Survey (JVAS), and 
a similar, complete radio sample, JVAS++, 
using the APM scans of the POSS-I plates.
It yields an identification rate of 83\%, with a completeness and 
reliability of both 99\%. About 20\% is identified with extended 
objects, eg. galaxies. 
The identification rate appears to drop towards lower flux densities,
and towards steeper radio spectra, especially for the stellar classifications.
Furthermore, the optical fluxes of  
quasars with faint radio flux densities appear to be biased towards 
fainter magnitudes, although there is a large spread in the
optical-to-radio spectral index.
It is shown that this large spread in  radio-to-optical 
spectral index may be caused by source to source variations
in the Doppler boosting of the synchrotron emission.

\section*{Acknowledgements}
We thank Robert Sharp for calculating the optical colours of a 
template quasar as function of redshift. We thank the referee,
Jasper Wall, for valuable comments.
We acknowledge the use of NASA's SkyView facility
(http://skyview.gsfc.nasa.gov) located at NASA Goddard
Space Flight Center.
This research was in part funded by the European Commission under
contract ERBFMRX-CT96-0034 (CERES).

{}

\section*{Appendix A}
Tables \ref{table1},\ref{table2}, \ref{table3}  show the optical-radio
catalogue for the source lists I, II, and III, as defined in section 2.
For each table, column 1 gives the name (JVS=JVAS, CLS=CLASS), 
columns 2 the radio position (J2000),
columns 3, 4, and 5 give the optical-radio offset in right ascension, 
declination and in total. Column 6 gives the logarithm of the 
likelihood ratio. Columns 7 to 12 give the APM magnitude, 
classification (-1=stellar, 1=extended, 2=merged) 
and psf for the red and blue plate respectively. 
Column 13 gives the APM 
colour, columns 14, to 17 give the radio flux densities at 5 GHz (GB6), 
1.4 GHz (NVSS and GB1400), and at 8.4 GHz (JVAS). 
Column 18 gives the redshift as found 
in the NASA/IPAC Extragalactic Database (NED). Column 19 gives a possible 
comment. The comments are explained in table 4.
The optical parameters are not shown if $\Delta\alpha , \Delta\delta >3.0''$,
unless the check by eye has shown that a larger optical-radio position offset
is the result of a bright galaxy or a merged object identification, or due to 
extended radio emission. 

Figure \ref{cases} show all the sources for which the APM did not give
a good representation. The three panels show $4'\times4'$ 
representations of the Digitized Sky Survey, The APM-red, and 
the APM blue data. The number in the left corner 
indicates the optical-radio position offset.

Figure \ref{extended} shows contour plots of NVSS data of objects in the 
complete sample 
showing extended structure on arcminute scale. The greyscales represent
optical Digitized Sky Survey data. Image sizes are 12'. The dashed 
circle indicate a 70'' radius around the GB6 position.

\begin{figure*}
\vbox to220mm{\vfil Figure 16 (gif) to go here}
\caption{ Extended radio sources in the complete sample. \label{extended}}
\end{figure*}
\begin{figure*}
\vbox to220mm{\vfil Figure 17 (gif) to go here}
\caption{\label{cases} All the JVAS sources with anomalies in the APM data.}
\end{figure*}

\clearpage

\begin{table*}
\vbox to220mm{\vfil table 2 can be found at www.roe.ac.uk/~ignas}
\caption{}
%\vfil}
\label{table1}
\end{table*}

\clearpage

\begin{table*}
\vbox to220mm{\vfil table 3 can be found at www.roe.ac.uk/~ignas}
\caption{}
%\vfil}
\label{table2}
\end{table*}
\clearpage

\begin{table*}
\vbox to220mm{\vfil table 4 can be found at www.roe.ac.uk/~ignas}
\caption{}
%\vfil}
\label{table3}
\end{table*}
\clearpage

\begin{table}
\caption{Explanation of the comments in tables 2, 3, and 4.}
\begin{tabular}{cl}
Code& Comment\\
a&Blended Object\\
b&Bright Galaxy\\
c&Possible ID on DSS?\\
d&Group/Cluster?\\
e&Bright Star nearby\\
f&ID not Possible\\
g&wrong representation in APM\\
h&Possible Blended Object?\\
i&ID is Correct\\
j&Offset in Red Plate\\
k&Blended Object in Blue\\
l&Blended Object in Red\\
m&Red and Blue APM are shifted\\
n&Bright Galaxy Nearby\\
o&DSS not Uniform  \\
p&NVSS extended\\
q&NVSS several components\\
r&NVSS Wide angle tail\\
s&NVSS Double\\
t&NVSS triple\\
u&Radio Source is Lobe\\
\end{tabular}
\end{table}
\end{document}